\documentstyle[12pt,lathuile,epsf]{article}
\begin{document}
\thispagestyle{empty}
\begin{flushright}
{CERN-TH/2001-195}\\
hep-ph/0107247
\end{flushright}
\vskip 3.0 true cm 

\begin{center}
{\Large \bf Neutrino deep-inelastic scattering:\\
new experimental and theoretical results}\\
[25 pt]
{\bf A. L. Kataev}         \\
{\em Theoretical Physics Division, CERN CH-1211 Geneva, Switzerland} \\
{\em and Institute for Nuclear Research of the Academy of Sciences 
of Russia,}\\
{\em 117312, Moscow, Russia} \\

%\maketitle
\baselineskip=14.5pt
\vskip 1.0 true cm  
{\bf Abstract}
\end{center}
A review of  recent experimental and theoretical studies of 
characteristics of neutrino deep-inelastic scattering is presented.
Special attention is paid to the  determination of $\alpha_s$ and $1/Q^2$
non-perturbative effects from the  QCD fits to $xF_3$ data at different 
orders of perturbation theory, with the help of  
several theoretical methods.  
%\end{abstract}
%
\vskip 2cm
\begin{center}
Contributed to the Proceedings of   Les Rencontres de la Vall\'ee d'Aoste,\\
La Thuile, Italy, March 4-10, 2001
\end{center}
\vskip 2cm

\begin{flushleft}
{CERN-TH/2001-195}\\
July 2001
\end{flushleft}

\newpage
\section{Introduction}
The neutrino deep-inelastic scattering (DIS) is continuing to serve
as  the 
classical tool for probing the nucleon  structure 
(for  reviews, see 
e.g. Refs.\cite{Conrad:1998ne}\cite{Kataev:1994ty}). In the last few 
years some progress was made 
in more detailed experimental and theoretical studies 
of the behaviour of the cross-sections of $\nu N$ DIS, and 
in the extraction of the  structure 
functions (SFs)  $xF_3$ and $F_2$.
Among the data that are currently under active analysis 
are the ones provided by the CCFR/NuTeV 
collaboration at the Fermilab Tevatron (see e.g. 
Refs.\cite{Seligman:1997mc}\cite{Yang:2001ju}), the experimental results of theJINR--IHEP Neutrino Detector collaboration\cite{Alekhin:2001zj}  collected some time 
ago at the  IHEP (Protvino) U-70 proton synchrotron,  
and the preliminary data of the  CHORUS collaboration,  obtained recently   
at the CERN   SPS\cite{Oldeman:1999pe}\cite{Oldeman:2000}.

\begin{figure}[t]
\begin{center}
\vspace{9.0cm}
\includegraphics{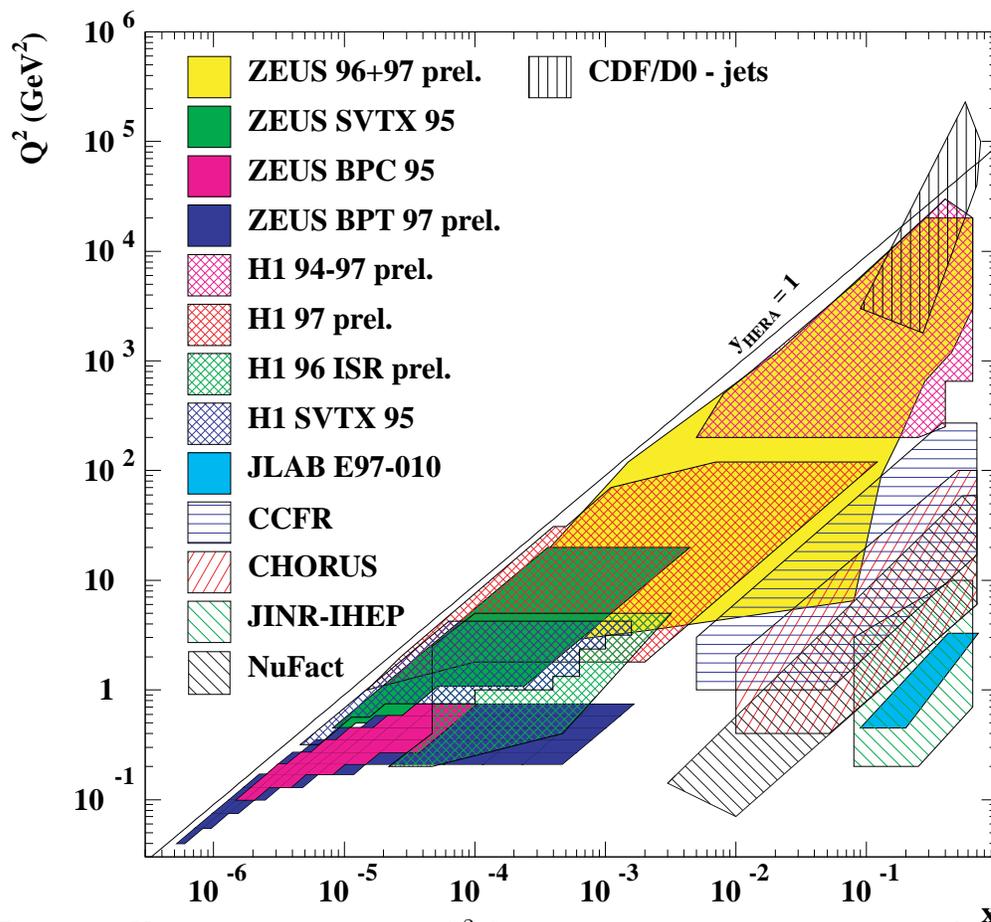}
%\end{center}
%\end{figure} 
\caption{\it 
      Kinematic regions in $x$-$Q^2$ for cross-section  
    measurements in deep inelastic $ep$ scattering,
     $\nu$ scattering and for triple differential
    jet cross-section measurements in $p\bar{p}$ collisions 
(from Ref.\cite{DeRujula}).
    \label{xfig} }
\end{center}
\end{figure}

The kinematical regions currently available  for cross-section measurements 
are shown on   the plot of Fig.~\ref{xfig}, taken 
from  Ref.\cite{Bassler:1999bv}. 
This figure clearly shows that the  
above-mentioned     
three experiments were performed in different kinematical 
regions, which overlap in part only. Thus they 
provide complementary information 
about the  behaviour of SFs in different  regimes.  
Moreover,  additional more precise data for $\nu N$ 
DIS cross-sections can be obtained in the  future at neutrino factories. 
If the energy of the   neutrino beam is fixed at  
$E_{\nu}=50~{\rm GeV}$, 
experiments should penetrate into  
the physical region that was  added to  Fig.1 in  Ref.\cite{DeRujula}.
This region overlaps in part with  those where  the 
above-mentioned three experimental collaborations were working. 
Therefore,
 the studies  of the  
data  on $\nu N$ DIS characteristics available at present  can 
be important milestones in the  planning of more precise DIS experiments 
at neutrino factories
\cite{Mangano:2001mj}.

\section{Discussions of some new  experimental results}

Recent  interesting  experimental news came from the  model-independent 
re-extraction of the behaviour of the $F_2$ $\nu N$ SF\cite{Yang:2001ju} 
from the 
CCFR'97 data\cite{Seligman:1997mc}. 
The re-analysis of Ref.\cite{Yang:2001ju}, which does 
not affect previous CCFR'97 $xF_3$ results, removed the 
widely discussed discrepancy that existed at  $x<0.1$ between the behaviour of CCFR'97\\ 
$F_2$ \cite{Seligman:1997mc} and that  obtained by the  NMC collaboration\cite{Amaudruz:1995tq} from the process of $\mu N$ DIS. 
In addition to the new extraction of $F_2$ from the differential 
cross-sections of CCFR, the first measurement of 
$\Delta xF_3=xF_3^{\nu}-xF_3^{\overline{\nu}}$ was also 
performed\cite{Yang:2001ju}. However,  it is important to stress
that none of the  considered 
non-perturbative, perturbative, and theoretical effects, including 
charm production,  
are still unable to describe properly the  experimental 
results obtained for  $\Delta xF_3$ \cite{Kretzer:2001mb}.

The  information about the corrected behaviour of CCFR $F_2$  
may    be really important for the 
continuation of the work of the  CHORUS collaboration. Indeed, 
in the kinematical region where both sets of $\nu N$ DIS data overlap,
the preliminary 
CHORUS results  for $xF_3$ and $F_2$\cite{Oldeman:2000} 
agree with the ones provided by the  CCFR collaboration in 
1997\cite{Seligman:1997mc}. 
Therefore, the preliminary CHORUS $F_2$ data should show 
a pattern identical to that found in the CCFR'97 analysis\cite{Seligman:1997mc},
i.e.  exceeding by 10--15$\%$ $F_2$ NMC measurements at $x<0.1$.
This excess is beyond the existing statistical and systematic 
errors of discussed DIS experiments.   
Moreover, the inclusion of these wrong  
CCFR'97 $F_2$ points into the next-to-leading order (NLO)  QCD  fits, 
performed  with the help of the DGLAP method\cite{DGLAP}, 
leads  to the erroneous low-$x$ 
behaviour of the gluon distribution 
$xG(x,9~{\rm GeV}^2)\sim x^{b_G}$ with $b_G=0.0092\pm 0.0073$ 
\cite{Alekhin:1999df}. It  is in evident contradiction with 
the number, obtained previously  from the NLO combined analysis of the 
data from  HERA 
and the CERN SPS\cite{Alekhin:1999za}, namely 
 $b_G=-0.267\pm 0.043$ at $Q_0^2=9~{\rm GeV}^2$.
Taking into account 
new CCFR model-independent 
extractions of $F_2$\cite{Yang:2001ju}, 
it seems worth while to perform  
more careful studies  of the preliminary  CHORUS  results.  
Moreover, it is  rather 
interesting to try to  verify from the CHORUS data  the  experimental 
behaviour  of  $\Delta xF_3$, found in  Ref.\cite{Yang:2001ju}.   

\begin{figure}[t]
\begin{center}
\vspace{5.0cm}
\includegraphics{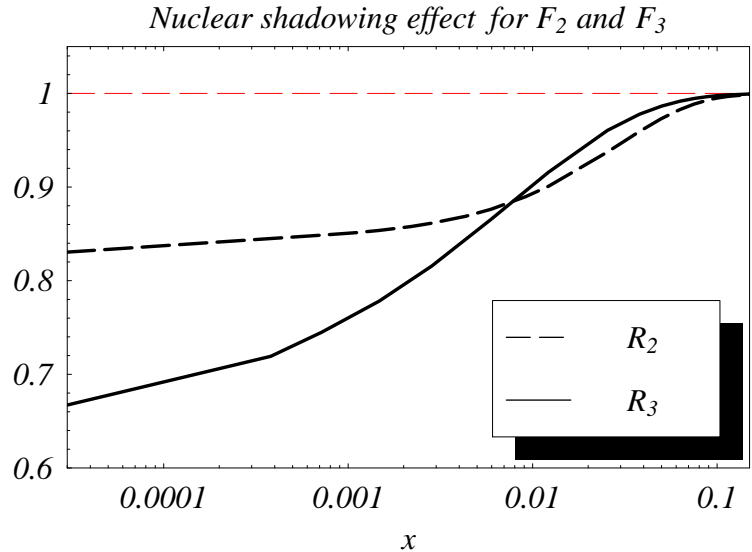}

%\end{center}
%\end{figure} 
\caption{\it 
      The ratios of a heavy target to the free nucleon SFs 
   $R_2=F_2^{A}/F_2^{N}$ and $R_3=F_3^{A}/F_3^{N}$ 
calculated for $^{56}${\rm Fe} nucleus in the region of small 
$x$ and $Q^2=10~{\rm GeV}^2$.
(the figure of Kulagin from Ref.\cite{Mangano:2001mj})    
\label{nuclfig} }
\end{center}
\end{figure}

It should be stressed that the CHORUS experiment has an  
attractive feature. Indeed, as can be clearly seen from Fig.1, 
it provides information about $\nu N$ DIS SFs 
in the region of rather low $Q^2$ 
and low $x$, which complement in part the one where the  CCFR'97 data 
were extracted. In this kinematical 
domain, theoretical contributions of $1/Q^2$ and 
nuclear corrections can  play an important role. Leaving 
for a while the discussions of power-suppressed terms, we  stress 
that  the CHORUS collaboration was using a  lead target, while the 
CCFR target 
is made of iron. 
Possibilities are therefore really open to study nuclear effects in 
neutrino DIS; as shown in calculations reported in Ref.\cite{Kulagin:1998wc},
these effects  can be of great  importance. 
A comparison of these effects for the  cases of $F_2$ and $xF_3$   
neutrino DIS SFs  is depicted in Fig.~\ref{nuclfig}, constructed 
for the detailed work of  Ref.\cite{Mangano:2001mj}.

Another interesting possibility  of  DIS experiments is   the extraction, 
from their chraracteristics, 
of 
non-perturbative  power-suppressed terms 
and the values 
of $\alpha_s(M_Z)$. 
This question was considered in 
Refs.\cite{Kataev:1998nc}$^-$\cite{Kataev:2001kk}  in the 
process of  
the next-to-next-to-leading order (NNLO) 
QCD fits to different data,   
and in Ref.\cite{Alekhin:1999hy}, while performing NLO fits 
to the experimental results for charged-leptons DIS SFs.  
The    
most recent 
outcomes  of the NNLO  analysis  of the   
CCFR'97 
$xF_3$ data\cite{Kataev:2001kk} will be discussed 
in the  next section. Here it is worth while emphasizing that 
the  works of Refs.\cite{Kataev:1998nc}$^-$\cite{Martin:2000gq}
\cite{Schaefer:2001uh}\cite{Kataev:2001kk}
agree in their conclusion that the inclusion of the 
NNLO QCD corrections into the fits has a tendency 
to decrease the extracted values 
of non-perturbative $1/Q^2$-terms. Whether this is a general theoretical 
feature (see e.g. Refs.\cite{Penin:1997zk}\cite{Dokshitzer:1999ai}) 
or it is related to the lack  of precision of 
the analysed data might be clarified in the future, if the ideas of more 
detailed experiments on  
neutrino DIS at  neutrino factories\cite{Mangano:2001mj} are 
realized.

It should be noted that the situation  at  NLO  is 
more transparent. Indeed, 
the Jacobi polynomial fits  of Refs.\cite{Kataev:1998nc}
\cite{Kataev:2000bp}\cite{Kataev:2001kk} 
demonstrated 
that  it is then possible to determine both $1/Q^2$ twist-4 effects 
and $\alpha_s(M_Z)$ from the  CCFR'97 $xF_3$ experimental results, 
cut at $Q^2\geq 5~{\rm GeV}^2$. Moreover, 
the NLO DGLAP fits of Ref.\cite{Alekhin:1999df} confirmed this feature, 
using both $xF_3$ and a cut at $x>0.1$ on  $F_2$ $\nu N$ CCFR'97   
data. A similar conclusion was made in the process 
of NLO DGLAP fits of intermediate-energy charged leptons data
\cite{Alekhin:1999hy}. 
However, it is obvious that it is simpler to detect non-perturbative 
effects in a smaller $Q^2$-region.

The most recent example of such a  determination was given by  
the NLO DGLAP analysis\cite{Alekhin:2001zj}
of  DIS neutrino data from  JINR--IHEP Neutrino Detector collaboration,
which collected rather large statistics 
(5987 neutrino and 741 antineutrino charged-current events)
in a rather low-energy region $0.55~{\rm GeV}^2\leq 
Q^2\leq 20~{\rm GeV}^2$. The non-perturbative $1/Q^2$ correction 
to the perturbation theory (PT)  behaviour of   $xF_3$  was parametrized using the infrared renormalon (IRR) model 
of Ref.\cite{Dasgupta:1996hh}, namely    
\begin{equation}
xF_3(x,Q^2)= xF_3^{PT}(x,Q^2)+\frac{h(x)}{Q^2}~~~~,
\end{equation}
where 
\begin{equation}
h(x)=A_2^{'}\int_x^1\frac{dz}{z}C_3^{IRR}(z)p_{NS}(x/z,Q^2)~~~~,
\end{equation}
$C_3^{IRR}$ is the IRR model coefficient function, 
$xp_{NS}(x,Q_0^2)=Ax^{b_{NS}}(1-x)^{c_{NS}}$ 
is the boundary parton distribution, defined  at $Q_0^2=0.5~{\rm GeV}^2$
at NLO, 
and $A_2^{'}$ is the normalization  parameter, which should be determined  
from the fits. It is sometimes expressed through the 
parameter $\Lambda_3$ as 
\begin{equation}
A_2^{'}=-\frac{2C_F}{\beta_0}\Lambda_3^2~~~,
\end{equation}
where $C_F=4/3$ and $\beta_0=(11-2/3f)$ is the first 
coefficient of the QCD $\beta$-function. Fixing 
$\alpha_s(M_Z)=0.118$, which corresponds to its  world-average value, 
$\Lambda_3$ was extracted  
from the fits to the  IHEP--JINR Neutrino 
Dectector  and CCFR'97 data\cite{Alekhin:2001zj}. The results  are 
presented in Table 1, taken from Ref.\cite{Alekhin:2001zj}.
\begin{table}
\centering
\caption{ \it The results of the fit, in  Ref.\cite{Alekhin:2001zj}, 
of the IRR model 
to the data from different neutrino experiments. The 
value of $\chi^2$ over the number of points (np)  is given.
}
\vskip 0.1 in
\begin{tabular}{|l|c|c|} \hline
Experiment          &  $\Lambda_3^2$[GeV$^2$] & $\chi^2/np$ \\
\hline
\hline
IHEP--JINR    & 0.69 $\pm$ 0.37    &  3/12                    \\ \hline 
CCFR'97     & 0.36 $\pm$0.22     &  253/222 \\
\hline
\end{tabular}
\label{extab}
\end{table}
Note that 
the two  expressions for $\Lambda_3^2$ from Table 1 are comparable 
within the  errors.
Averaging the numbers  for $\Lambda_3^2$  and transforming them to  $A_2^{'}$, 
the authors of Ref.\cite{Alekhin:2001zj} obtained the following value: 
\begin{equation}
A_2^{'}=-0.130 \pm 0.056~({\rm exp})~~{\rm GeV}^2~~~,
\end{equation}
where the  error includes   both statistical and systematic
experimental uncertainties. It is in agreement with 
the value extracted 
from  
the NLO Jacobi polynomial analysis of the CCFR'97 $xF_3$ behaviour,  
cut at $Q^2\geq 5~{\rm  GeV}^2$ \cite{Kataev:1998nc}
\cite{Kataev:2000bp}
\cite{Kataev:2001kk}. Indeed, at NLO, the most detailed fits  
of Ref.\cite{Kataev:2001kk}
give : 
\begin{equation}
A_2^{'}=-0.125\pm 0.053~~({\rm stat})~~~{\rm GeV}^2~~~.
\end{equation}    
Note, however, that the error in Eq.(5) does not include 
the systematic uncertainties. Therefore, Eq.(4) 
is  the most precise up-to-date value of the IRR model parameter $A_2^{'}$.

\section{New QCD fits to CCFR $xF_3$ data: NNLO  and 
beyond}

We can start  the discussion on the  phenomenological 
application of some new N$^3$LO 
perturbative QCD results on the coefficient functions of odd  moments 
of $xF_3$ and on the NNLO 
approximations   for the related anomalous dimensions\cite{Retey:2001nq}
(which are complementary  to those obtained in Ref.\cite{Larin:1994vu}
in the case of even moments 
for the  $F_2$ SF 
of charged-leptons DIS),  
in combination with  the  
NNLO expressions  for  the coefficient 
functions\cite{Zijlstra:1992kj} that were recently confirmed  in  
Ref.\cite{Moch:2000eb}.
\subsection{The application of the Jacobi polynomial method}
It is appropriate, at this point to  
recall the basic ideas  of the Jacobi polynomial method\cite{Parisi:1979jv} 
which was developed in Refs.\cite{Chyla:1986eb}\cite{Krivokhizhin:1987rz} 
and  
was previously  used in the analysis of the BCDMS charged-leptons    
DIS data at  NLO\cite{Benvenuti:1987zm}, 
and in the non-singlet approximation at NNLO\cite{Parente:1994bf}. 
In the case of the  analysis of the  CCFR 
$xF_3$ data, the Jacobi polynomial method was applied  at 
NLO in Refs.\cite{Kataev:1994rj}\cite{Chyla:1995bt}  and proved 
useful for performing fits at the NNLO 
\cite{Kataev:1998nc}\cite{Kataev:2000bp}\cite{Kataev:2001kk}  
and approximate N$^3$LO levels, 
\cite{Kataev:2000bp}\cite{Kataev:2001kk} with and without 
twist-4 corrections (see discussion below).

This method allows the 
reconstruction of the SF (say $xF_3$) from the {\bf finite} number 
of Mellin moments, namely 
\begin{equation}
xF_3^{N_{max}}(x,Q^2)=x^{\alpha}(1-x)^{\beta}\sum_{n=0}^{N_{max}}
\Theta_n^{\alpha\beta}(x)\sum_{j=0}^{n}c_j^{(n)}(\alpha,\beta)M^{TMC}_{j+2,F_3}(Q^2) +\frac{h(x)}{Q^2}~~~~,
\end{equation}
where $\Theta_n^{\alpha,\beta}$ are the Jacobi orthogonal polynomials
with parameters $\alpha$, $\beta$; $c_j^{(n)}(\alpha,\beta)$ is the combination of Euler $\Gamma$-functions.
It  increases factorially  with increasing $n$.  The  Mellin moments 
$M_{n,F_3}^{TMC}$ include information on the $1/Q^2$ target mass 
corrections and are  defined as 
\begin{equation}
M_{n,F_3}^{TMC}(Q^2)=\int_0^1x^{n-1}F_3^{PT}(x,Q^2)dx+\frac{n(n+1)}{n+2}
\frac{M^2_{nucl}}{Q^2}M_{n+2,F_3}(Q^2)~~~.
\end{equation}  
The contribution  of the twist-4 terms to Eq.(6) is parametrized 
with the help of 
the function $h(x)$. It  will be  neglected  
for our first stage of discussions.  

Fixing now the behaviour  $xF_3$ at the initial scale $Q_0^2$ as 
\begin{equation}
xF_3^{PT}(x,Q_0^2)=A(Q_0^2)x^{b(Q_0^2)}(1-x)^{c(Q_0^2)}(1+\gamma(Q_0^2))~~~~,
\end{equation}
calculating the related Mellin moments and transforming  
them to  experimentally accessible regions 
with the help of the renormalization group technique at  
LO, NLO, NNLO and approximate N$^3$LO 
(the explicit  formulae for the renormalization group evolution can be 
found in Ref.\cite{Kataev:2001kk}), substituting the 
renormalization-group-improved expression for $M_{n,F_3}^{TMC}(Q^2)$ 
into Eq.~(6), and 
performing 
the fits to the experimental data, it is possible  to determine 5 parameters, 
namely $A,b,c,\gamma$ and $\Lambda_{\overline{MS}}^{(4)}$; this  enters 
the QCD coupling constant $\alpha_s$, defined up to N$^3$LO,  
by means of the solution 
of the renormalization group equation for  
 the explicitly known  4-loop approximation  of the 
QCD $\beta$-function\cite{vanRitbergen:1997va}.
\begin{table}
\centering
\caption{ \it The NNLO    
results  of the parameters $A,b,c$ of the model 
for $xF_3$ determined, 
in Ref.\cite{Kataev:2001kk} and their  comparison with the values obtained     
in  Ref.\cite{Kataev:2000dp}. The new ones are marked 
by bold type.} 
\vskip 0.1 in
\begin{tabular}{||c|c|c|c|c|c||}
\hline
Order/$N_{max}$ & $Q_0^2$&  {\it A}  & {\it b} & {\it c} & 
$\chi^2$/np \\ \hline 
NNLO/6 & 5 GeV$^2$ 
   &  4.25$\pm 0.38$ &  0.66$\pm$0.03 &  
3.56$\pm$0.07 
 &  78.4/86  \\

{\bf NNLO/9} &
   & {\bf 3.73$\pm$0.68} & {\bf 0.63$\pm$0.05} & {\bf 
3.52$\pm$0.08}
 & {\bf 72.4/86}  \\

%&8 GeV$^2$ 
%  & {\bf 332$\pm$35} & {\bf 4.10$\pm$0.39} & {\bf 0.63$\pm$0.03} & {\bf 
%3.66$\pm$0.06} & {\bf 1.33$\pm$0.33}
% & {\bf 73.6/86}  \\
%&8 GeV$^2$ 
%  & {\it 312$\pm$33} & {\it 4.42$\pm$0.36} & {\it 0.66$\pm$0.03} & {\it 
%3.68$\pm$0.07} & {\it 1.13$\pm$0.31}
% & {\it  76.5/86}  \\
NNLO /6 &10 GeV$^2$ 
   &  4.50$\pm$0.36 &  0.65$\pm$0.03 &  
3.73$\pm$0.07 & 
  76.3/86  \\
{\bf NNLO/9}  &
   & {\bf 4.21$\pm$0.35} & {\bf 0.63$\pm$0.03} & {\bf 
3.73$\pm$0.07} & 
 {\bf 74.2/86}  \\
%&12.59 GeV$^2$
%  & 333$\pm$35 & 4.31$\pm$0.43 & 0.63$\pm$0.03 & 3.79$\pm$0.06 & 1.12$\pm$0.35
% & 74.7/86  \\

NNLO/6 &20 GeV$^2$ 
   &  4.70$\pm$0.34 &  0.65$\pm$0.03 &  
3.88$\pm$0.08 
 &  77.0/86  \\
{\bf NNLO/9} &
   & {\bf 4.49$\pm$0.25} & {\bf 0.63$\pm$0.02} & {\bf 
3.89$\pm$0.06} 
 & {\bf 75.8/86}  \\

NNLO/6 &100 GeV$^2$ 
   &  4.91$\pm$0.28 &  0.63$\pm$0.02 &  
4.11$\pm$0.10 
 &  80.0/86  \\
{\bf NNLO/9} &
  & {\bf 4.74$\pm$0.32} & {\bf 0.61$\pm$0.02} & {\bf 
4.14$\pm$0.09} 
 & {\bf 77.8/86}  \\
\hline
\end{tabular}
\label{partab}
\end{table}
In Table 2 the new  NNLO results of Ref.\cite{Kataev:2001kk} for the 
parameters $A,b,c$ of the model of Eq.(8) are presented. 
One can notice that, although these results agree with those 
of Ref.\cite{Kataev:2000dp} within the statistical errors,
the central values of the  new numbers are 
over 0.03 lower,  and the new fits have  smaller values of $\chi^2$.
  
It should be stressed  that the  new multiloop 
calculations of Ref.\cite{Retey:2001nq}  
allow us  to use more 
moments in Eq.~(6), namely $n=13$, which corresponds 
to fixing  $N_{max}=9$. All this information was effectively used 
in Ref.\cite{Kataev:2001kk}. 
Note that the previous, similar $xF_3$  fits 
of Refs.\cite{Kataev:1998nc}\cite{Kataev:2000bp} were made  in the case 
of $n=10$ and  $N_{max}=6$; they were  using the  approximate 
NNLO  expressions 
for non-singlet anomalous dimensions,  obtained from exact  
NNLO expressions    
for the   anomalous dimensions of 
even non-singlet moments of $F_2$, calculated 
in Ref.\cite{Larin:1994vu}. 
Thus the considerations of  Ref.\cite{Kataev:2001kk}
contain less uncertainties than the   
previous analysis of  Refs.\cite{Kataev:1998nc}\cite{Kataev:2000bp}.
The NNLO and N$^3$LO results for 
$\Lambda_{\overline{MS}}^{(4)}$, obtained in Ref.\cite{Kataev:2001kk}
in the process of twist-4 independent fits, are presented in Table 3. 
One can see that the application 
of new information from  Ref.\cite{Retey:2001nq}, which allowed  
$N_{max}$ in Eq. (6) to go from 6 to 9, 
leads  to better stability of 
$\Lambda_{\overline{MS}}^{(4)}$ with respect to changes of 
the initial scale $Q_0^2$,
and decreases the values of $\chi^2$.
\begin{table}
\centering
\caption{\it The $Q_0^2$ and $N_{max}$ 
dependence of $\Lambda_{\overline{MS}}^{(4)}$
( in  MeV)   from  Ref.\cite{Kataev:2001kk}. The values of $\chi^2$ are presented in 
parenthesis.}
\vskip 0.1 in
\begin{tabular}{||c|c|c|c|c|c|c|c|}
\hline
$N_{max}$& $Q_0^2$ (${\rm GeV}^2$) & 5 & 8 & 10 & 20 & 50 & 100 \\
%\hline
\hline
6& NNLO & 297$\pm$30 & 314$\pm$34 & 320$\pm$34 & 327$\pm$36 & 327$\pm$35 &
326$\pm$35\\
 &      &  (77.9)    &   (76.3)   &  (76.2)    &  (76.9)    &  (78.5)   &
(79.5) \\
%\hline

7& NNLO & 326$\pm$34 & 327$\pm$35 & 327$\pm$35 & 326$\pm$36 & 327$\pm$36 &
328$\pm$35\\
 &      &  (75.9)    &   (76.7)   &  (77.1)    &  (78.1)    &  (78.8)   &
(78.7) \\
%\hline

8& NNLO & 334$\pm$35 & 334$\pm$35 & 333$\pm$35 & 331$\pm$35 & 328$\pm$35 &
328$\pm$35\\
 &      &  (74.3)    &   (75.7)   &  (76.2)    &  (77.4)    &  (78.3)   &
(78.5) \\
%\hline

9& NNLO & 330$\pm$33 & 332$\pm$35 & 333$\pm$34 & 331$\pm$37 & 330$\pm$35 &
329$\pm$35\\
 &      &  (72.4)    &   (73.6)   &  (74.7)    &  (75.8)    &  (76.7)   &
(77.8) \\
\hline

%NNLO$^*$ & 284$\pm$28 & 312$\pm$33 & 318$\pm$33 & 326$\pm$35 & 326$\pm$36 &
%325$\pm$36 \\

6& N$^3$LO  & 303$\pm$29 & 317$\pm$31 & 321$\pm$32 & 325$\pm$33 & 325$\pm$
33 & 324$\pm$33 \\
 &          &  (76.4)    &  (75.6)    &  (75.7)    &  (76.6)    & (78.0) &
(78.7) \\ 
%\hline

7& N$^3$LO  & 328$\pm$32 & 326$\pm$33 & 325$\pm$33 & 322$\pm$33 & 324$\pm$
33 & 324$\pm$33 \\
 &          &  (76.2)    &  (77.0)    &  (77.3)    &  (78.2)    & (78.5) &
(78.2) \\

8& N$^3$LO  & 334$\pm$33 & 329$\pm$33 & 327$\pm$34 & 324$\pm$34 & 323$\pm$
34 & 324$\pm$34 \\
 &          &  (74.8)    &  (76.2)    &  (76.6)    &  (77.4)    & (77.3) &
(77.2) \\ 

9& N$^3$LO  & 330$\pm$31 & 329$\pm$34 & 329$\pm$32 & 325$\pm$33 & 325$\pm$
32 & 325$\pm$33 \\
 &          &  (73.3)    &  (74.6)    &  (75.7)    &  (76.4)    & (76.7) &
(76.8) \\ 
\hline 
\end{tabular}
\label{Lamtab}
\end{table}

It is interesting to compare the NNLO results for $b(Q_0^2)$ from Table 2 , 
which are almost 
$Q_0^2$-independent, with the calculations 
of the small-$x$ asymptotic behaviour 
of non-singlet contributions to  $F_1$ and the  spin-dependent SF $g_1$
performed in Ref.\cite{Ermolaev:2001sg} in 
all orders of 1-loop expression for  $\alpha_s$, using  in part  
the approach developed in Ref.\cite{Kirschner:1983di}.
In the process of calculations of  Ref.\cite{Ermolaev:2001sg} the following 
1-loop formula for $\alpha_s$ was used 
\begin{eqnarray}
\alpha_s(s)&=&\frac{4\pi}{\beta_0\ln(-s/\Lambda^2)}=\frac{4\pi}
{\beta_0[ln(s/\Lambda^2)-i\pi]} \\ \nonumber 
      &=&\frac{4\pi}{\beta_0}\bigg[\frac{\ln(s/\Lambda^2)}
{\ln^2(s/\Lambda^2)+\pi^2}+\frac{i\pi}{\ln^2(s/\Lambda^2)+\pi^2}\bigg]~~,
\end{eqnarray} 
where  $\Lambda=\Lambda^{(f=3)}=0.1$. 
Note
that the idea of taking into account 
the effects of $\pi^2$-terms 
in the perturbative expansion parameter is not new. It was 
previously used  in a number of works on the subject 
\cite{Radyushkin:1996kg}$^-$\cite{Pivovarov:1992rh}
(for recent applications see in particular Ref.\cite{Shirkov:2001qv} 
and  
Ref.\cite{Broadhurst:2001yc} where  other related  works were   
 discussed as well). 

\begin{figure}
\begin{center}
\begin{picture}(320,220)
\put(0,20){
\epsfbox{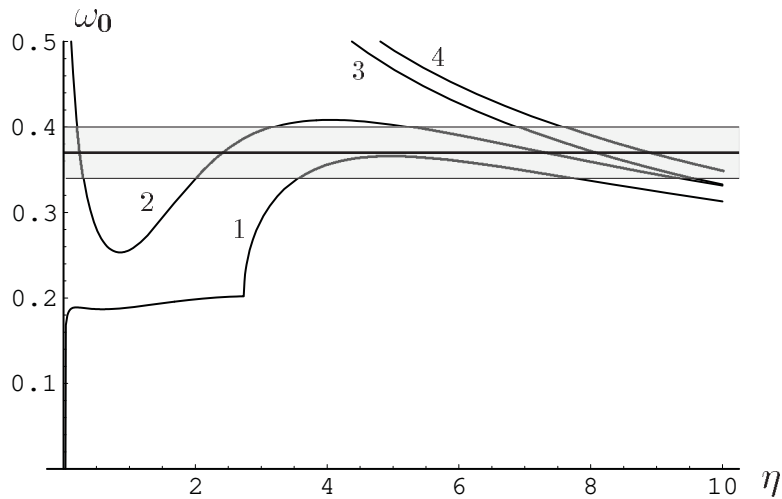}
}
\end{picture}
\end{center}
\caption{\it 
Dependence of  $\omega_0$ on 
$\eta=\ln(\mu^2/\Lambda^2)$. 1: for $F_1^{NS}$, 2: for
$g_1^{NS}$, 3 and 4: for $F_1^{NS}$ and $g_1^{NS}$, respectively,
without accounting for the $\pi^2$-terms (from Ref.\cite{Ermolaev:2001sg}).
The results of the fits of Ref.\cite{Kataev:2001kk} are added.}
\label{intercept}
\end{figure}
The solution of the  corresponding 
equations, which will not  be  presented  here, and the  application of 
Eq.~(9) allowed the authors of Ref.\cite{Ermolaev:2001sg} to 
obtain the following small-$x$ asymptotic behaviour of $F_1^{NS}$ 
and $g_1^{NS}$~: 
\begin{eqnarray}
F_1^{NS}&\sim& \bigg(\frac{1}{x}\bigg)^{\omega_0^{(+)}}
\bigg(\frac{Q^2}{\mu^2}\bigg)^
{\omega_0^{(+)}/2} \\ \nonumber 
g_1^{NS}&\sim& \bigg(\frac{1}{x}\bigg)^{\omega_0^{(-)}}
\bigg(\frac{Q^2}{\mu^2}\bigg)^
{\omega_0^{(-)}/2} \\ \nonumber~~~, 
\end{eqnarray}
where the powers $\omega_0^{(+)}$ and $\omega_0^{(-)}$ are 
drawn  in Fig. 3, taken from  Ref.\cite{Ermolaev:2001sg}.
This plot is supplemented with  the   
bounds on the values of $\omega_0=1-b(Q_0^2)$, which 
come from the results  presented in Table 2 
and 
obtained in the process of the fits to the  CCFR'97 data
for the non-singlet SF $xF_3$,  performed in Ref.\cite{Kataev:2001kk}
for  $Q_0^2\geq 5~ {\rm GeV}^2$.
These values  are lying in the area limited by straight lines, 
with the centre  at  $\omega_0=0.37$ shown in Fig. 3. 
It  is worth 
trying to  understand in more detail   the possible relations 
between the     
results 
of Ref.\cite{Ermolaev:2001sg} and those of 
 Ref.\cite{Kataev:2001kk}.
Note, in particular, that  the comparison seems to be 
more legitimate in the region $\eta\geq 8$, namely for 
$Q^2>\mu^2=30~{\rm GeV^2}$, than for low values of $\eta$, 
where  the 
$\pi^2$-effects are playing the dominant role. 
This region is not considerably affected even  by 
the transformation from $\Lambda^{(3)}\approx0.1~{\rm GeV}$, used 
in  Ref.\cite{Ermolaev:2001sg}, to $\Lambda^{(3)}\approx0.4~{\rm GeV}$, 
which follows  from  the $Q_0^2$-independent numbers 
of Table 3, obtained when   $N_{max}$ is fixed to 9 in Eq.~(6).

Consider now some other  results of the  
work of Ref.\cite{Kataev:2001kk}, and in particular  the extraction 
of twist-4 contributions and the value of  $\alpha_s(M_Z)$ at 
various orders of perturbation theory. To model the $1/Q^2$-term  
$h(x)$ in Eq.~(6),  
three  approaches were used in Ref.\cite{Kataev:2001kk}.
The first one is the 
IRR model  of Ref.\cite{Dasgupta:1996hh} 
(see Eq.~(2)).

After taking the Mellin moments from Eq.~(2) and applying 
the NNLO and N$^3$LO fits to the  CCFR'97  $xF_3$ data, the  
reductions of the  NLO value of $A_2^{'}$, presented in Eq.~(5), were observed.
At the NNLO and N$^3$LO the expressions for $A_2^{'}$  become comparable 
with zero, 
within the statistical errors\cite{Kataev:2001kk}, namely 
\begin{eqnarray}
{\rm NNLO}~~~~:~~~~A_2^{'}&=& -0.013\pm 0.051~~{\rm GeV}^2 \\ \nonumber 
{\rm N^3LO}~~~~:~~~~A_2^{'}&=& 0.038\pm 0.051~~{\rm GeV}^2~~~.
\end{eqnarray}
However, the  
related $\alpha_s(M_Z)$ results were determined in Ref.~\cite{Kataev:2001kk}
with 
reasonable errors: 
%\newpage
\begin{eqnarray}
{\rm NLO}~~~:~~~\alpha_s(M_Z)&=&0.120 \pm 0.002~({\rm stat})~\pm 
0.005~({\rm syst}) 
\\ \nonumber 
&&~~~~~~~~\pm 0.002~({\rm thresh})^{+0.010}_{-0.006}~({\rm scale})
\\ \nonumber  
{\rm NNLO}~~~:~~~\alpha_s(M_Z)&=&0.119 \pm 0.002~({\rm stat})~\pm 0.005~
({\rm syst}) 
\\ \nonumber 
&&~~~~~~~~~~\pm 0.002~({\rm thresh})^{+0.004}_{-0.002}~({\rm scale})
\\ \nonumber   
{\rm N^3LO}~~~:~~~\alpha_s(M_Z)&=&0.119 \pm 0.002~({\rm stat})~\pm 0.005~
({\rm syst}) 
\\ \nonumber 
&&~~~~~~~~~~\pm 0.002~({\rm thresh})^{+0.002}_{-0.001}~({\rm scale})
\end{eqnarray}
where the first theoretical uncertainty is due to the ambiguities of  
taking into account threshold effects 
while transforming the results for 
$\Lambda_{\overline{\rm MS}}^{(5)}$ to a world with $f=5$ numbers of 
active flavours 
(for a  detailed explanation of   how this uncertainty was fixed using the 
matching conditions of Ref.\cite{Chetyrkin:1997sg},  
see  Ref.\cite{Kataev:2001kk} and 
references  therein)
and the scale-dependence uncertainty was determined by  choosing 
the factorization and renormalization scales $\mu_F^2=\mu_R^2=
\mu_{\overline{MS}}^2~k$ and varying $k$ in the conventional interval 
$1/4\leq k \leq 4$. One can notice the drastic reduction of the 
scale-dependence uncertainties as a result of adding  
NNLO and N$^3$LO perturbative QCD corrections into the fits, 
tabulated in the case of $f=4$  
in Ref.\cite{Kataev:2001kk} (note that at the N$^3$LO the contributions to 
expanded  
anomalous-dimension terms  were modelled using [1/1] Pad\'e approximants).  
 
The results for $\alpha_s(M_Z)$, presented in Eq.~(12), should be 
compared with the ones obtained from the twist-4 independent 
Jacobi polynomial fits to  the 
CCFR'97 data at $N_{max}=9$\cite{Kataev:2001kk}, which give   
\begin{eqnarray}
{\rm NLO}~~~:~~~\alpha_s(M_Z)&=&0.118 \pm 0.002({\rm stat})
\pm 0.005({\rm syst}) \\ \nonumber
&&~~~~~\pm 0.002({\rm thresh})^{+0.007}_{-0.005}({\rm scale}) \\ \nonumber 
{\rm NNLO}~~~:~~~\alpha_s(M_Z)&=&0.119\pm 0.002({\rm stat})
\pm 0.005({\rm syst}) \\ \nonumber
&&~~~~~\pm 0.002({\rm thresh})^{+0.004}_{-0.002}({\rm scale}) \\ \nonumber
{\rm N^3LO}~~~:~~~\alpha_s(M_Z)&=&0.119\pm 0.002({\rm stat})
\pm 0.005({\rm syst}) \\ \nonumber
&&~~~~~\pm 0.002({\rm thresh})^{+0.002}_{-0.001}({\rm scale})
\end{eqnarray}
Notice that the effective minimization of the twist-4 contributions at the 
NNLO and N$^3$LO (see Eq.~(11)) is leading to  
rather closed  
NNLO and N$^3$LO values of  $\alpha_s(M_Z)$, which were  
obtained from  the fits with and without 
$1/Q^2$ corrections.  

It is worth   
stressing  that errors on  the scale dependence 
of the NLO and NNLO 
results from Eq.~(13) have definite support. Indeed, 
they are in agreement with the independent estimates   
\begin{equation}
\Delta\alpha_s(M_Z)_{NLO}=^{+0.006}_{-0.004}~~~,~~~ 
\Delta\alpha_s(M_Z)_{NNLO}=^{+0.0025}_{-0.0015}~~~,
\end{equation}
obtained in Ref.\cite{vanNeerven:2000ca}, which use the 
model constructed in this work 
for the 
NNLO NS DGLAP kernel.
 
In order to study  the second possibility of modelling  $1/Q^2$-effects 
using 
the parametrization of $h(x)$ by   
free constants  $h_{i}=h(x_i)$, where $x_i$ are the points in the 
experimental 
data binning, 9 parameters $h_i$ were used 
in Ref.\cite{Kataev:2001kk}. This choice distinguishes 
new fits from the ones performed in Refs.\cite{Kataev:1998nc}
\cite{Kataev:2000bp}, where 16 variables  $h_i$ were used.
The minimization of the number of free parameters 
was motivated by the works of 
Refs.\cite{Alekhin:1999hy}\cite{Alekhin:1999df}, where it was demonstrated 
that a decrease in the  number of fitted high-twist parameters 
decreases the correlation between their errors and make their extraction 
more reliable (the problems of  estimating theoretical uncertainties 
in the case of the choice of 16 free parameters $h_i$ were  also discussed 
in Ref.\cite{Forte:1998qq}). 
The choice of a smaller number of $h_i$ results 
in a more reliable description of the  $x$-shape of $h(x)$ for  the  
fits to the  CCFR $xF_3$ data. As in the process of the analogous  
fits of 
Refs.\cite{Kataev:1998nc}\cite{Kataev:2000bp}, the  
LO and NLO $x$-shapes of $h(x)$  obtained in 
Ref.\cite{Kataev:2001kk}
are  in  agreement with 
the prediction of the IRR model of  Ref.\cite{Dasgupta:1996hh}. 
The new  NNLO and N$^3$LO 
results  of Ref.\cite{Kataev:2001kk}, in agreement with 
the above-discussed tendency to an  overall minimization of the 
extracted contribution  of $h(x)$, reveal some 
new feature, namely an indication of an  
oscillating-type behaviour of $h(x)$ around $x=0$, 
albeit with rather small amplitude.  

The third model of $1/Q^2$-corrections, considered in Ref.\cite{Kataev:2001kk},
is directly  expressed in terms of Mellin moments, namely
\begin{equation}
M_n^{HT}(Q^2)=n\frac{B_2^{'}}{Q^2}M_n^{F_3}(Q^2)
\end{equation}
with the free parameter $B_2^{'}$. It is identical to the model used in 
Ref.\cite{Santiago:2001mh} for fixing theoretical uncertainties 
of the  extraction of $\alpha_s(M_Z)$ at  NNLO with the help of 
the Bernstein polynomial technique, which will be discussed below. 

In Ref.\cite{Kataev:2001kk} it was shown that the precision of the 
CCFR'97 
$xF_3$ data  allows a determination of   the value of $B_2^{'}$ together with 
$\alpha_s(M_Z)$ both at  LO and NLO. However, as  in the cases 
of the previous two models for $1/Q^2$-corrections, considered 
in Ref.\cite{Kataev:2001kk}, NNLO perturbative QCD 
effects   screen the contribution of 
non-perturbative $1/Q^2$-corrections, defined through Eq.~(15).
It should be recalled,  in these circumstances, that the    
 QCD fits of Ref.\cite{Abbott:1980as} to 
BEBC--Gargamelle\cite{Bosetti:1978kz}
and CDHS\cite{deGroot:1979hr} neutrino DIS data, performed   
over 20 years ago, 
did not allow a
discrimination between $1/Q^2$ and the  logarithmic description of scaling 
violation to be made. Therefore, it is possible to conclude that present 
neutrino 
DIS data now have become more precise. Indeed, 
their analysis shifted the  effect of perturbative 
screening of $1/Q^2$-corrections from LO to NNLO. The next generation 
of more detailed  tests of QCD in neutrino DIS is now on the 
agenda\cite{Mangano:2001mj}.

\subsection{The application of the Bernstein  polynomial method}

In this part of our  mini-review the basic steps of the Bernstein 
polynomial approach, proposed in Ref.\cite{Yndurain:1978wz} and recently 
used in the process of NNLO fits to the   CCFR'97 $xF_3$ data in 
Ref.\cite{Santiago:2001mh}, will be recalled.
The basic constructions of this approach are  the Bernstein averages 
for the $xF_3$ SF~:
\begin{equation}
F_{nk}^{F_3}(Q^2)=\int_0^1 dx p_{nk}(x)xF_3(x,Q^2)~~,
\end{equation}
where $p_{nk}(x)$ are the Bernstein polynomials, which can be presented,  
when $k\leq n$, in  the following form:
\begin{equation}
p_{nk}(x)= p(n,k)\sum_{l=0}^{n-k}\frac{(-1)^l}{ l!(n-k-l)!}x^{2(k+l)+1}~~~,
\end{equation}
where $p(n,k)$ is defined as (see e.g.\cite{Santiago:2001mh}):
\begin{equation}
p(n,k)=
\frac{2(n-k)!\Gamma(n+\frac{3}{2})}{\Gamma(k+\frac{1}{2})\Gamma(n-k+1)}~~.
\end{equation}
Using Eqs. (16)--(18), it is possible to express the Bernstein averages 
for $xF_3$  through $xF_3$  odd  Mellin moments as: 
\begin{equation}
F_{nk}^{F_3}(Q^2)=p(n,k)\sum_{l=0}^{n-k}\frac{(-1)^l}{l!(n-k-l)!} M_{(2k+2l+1),F_3}(Q^2)~~.
\end{equation}

The next step is similar to the one used in the process of applications 
of the Jacobi polynomial technique. 
At some 
initial scale $Q_0^2$, $xF_3$ can be parametrized through Eq.(8), 
several   odd Mellin moments of
$xF_3$ at $Q_0^2$ defined and then transformed  using NNLO  
renormalization group equations 
at the  appropriate  values of $Q^2$, 
which enter into the kinematical region of the analysed experimental data. 
Forming now Bernstein averages of Eq.~(18) it is possible to fit them to 
their experimental values. In the case of the  CCFR'97 $xF_3$ data, 
fits were made  in the kinematical region 7.9 GeV$^2$~$\leq Q^2 \leq$ 
125.9 GeV$^2$ \cite{Santiago:2001mh}.  
The following numbers  for $\alpha_s(M_Z)$ 
were obtained\cite{Santiago:2001mh} from these fits :  
\begin{eqnarray}
{\rm NLO}~~~~:~~~~\alpha_s(M_Z)&=&0.116 \pm 0.004~({\rm exp}) \\ \nonumber 
{\rm NNLO}~~~:~~~~\alpha_s(M_Z)&=&0.1153 \pm 0.004~({\rm exp})~~,
\end{eqnarray}
The final NNLO expression, which includes the estimates of some theoretical 
uncertainties, is \cite{Santiago:2001mh}:
\begin{equation}
{\rm NNLO}~~:~~~~\alpha_s(M_Z)= 0.1153 \pm 0.0041~({\rm exp})\pm 0.0061~
({\rm theor})~~~,
\end{equation}
It is worth while to mention that, despite the qualitative agreement, 
the central NLO values of Eq.(20), obtained with the help of
the  Bernstein polynomial technique,
are lower than the existing determinations of $\alpha_s(M_Z)$ 
from the CCFR'97 $xF_3$ data, which result from  the NLO DGLAP analysis
\cite{Seligman:1997mc}\cite{Alekhin:1999df}
and the application of the Jacobi polynomial technique\cite{Kataev:1998nc}
\cite{Kataev:2000bp}\cite{Kataev:2001kk}. 
Moreover, at NNLO, the result of Eq. (21) intersects with 
the  NNLO determination of $\alpha_s(M_Z)$ of Ref.\cite{Kataev:2001kk} 
(see Eqs.(12) and (13)) 
within existing errors only.
The comparison  between  the results of the 
Jacobi and Bernstein polynomial determinations  of  
$\alpha_s(M_Z)$ and of the related theoretical uncertainties was presented in  Ref.\cite{Kataev:2001kk}. 
In the process of these studies, 
definite  
disagreements were revealed 
between some  results of the works of Ref.\cite{Santiago:2001mh}
and Ref.\cite{Kataev:2001kk}.  
The origin of these disagreements   
is unclear at present and stimulates a  more detailed analysis  
of the NNLO  realizations  
of the Jacobi  and Bernstein  polynomial approaches. 
Note, however, that the  definite choice of 
the scale parameter in the Jacobi polynomial fits  
leads to improving the agreement 
of the results of applications of the two  methods\cite{Kataev:2001kk}. 
In view of this observation, 
it is possible that the results of Ref.\cite{Santiago:2001mh} 
contain larger 
theoretical 
uncertainties due to the neglect of scale-dependence ambiguities.
On the other hand, contrary to the Bernstein polynomial analysis, 
the NNLO Jacobi polynomial fits of Ref.\cite{Kataev:2001kk} also  
used 
approximate information about the values of the NNLO corrections to anomalous 
dimensions of even moments of $xF_3$. It should be stressed that this 
approximation can  be   eliminated after completing the 
program of explicit calculations
of NNLO contributions to non-singlet DGLAP kernels, which is 
now in progress\cite{Vermaseren}. 
As to the current applications of the  DGLAP method in the 
concrete NNLO fits to DIS data,
they can in principle   be based on the machinery of the 
Bayesian treatment of systematic 
errors of DIS data (see e.g. Ref.\cite{Alekhin:1999za}) and 
the approximate NNLO models 
of DGLAP kernels, constructed  in Refs.\cite{vanNeerven:2000ca}
\cite{vanNeerven:2000uj}\cite{vanNeerven:2000wp}.

\newpage
\begin{flushleft}
{\bf Acknowledgements}
\end{flushleft}
I am  grateful to  G. Parente and A. V. Sidorov for our long and fruitful  
collaboration,  which led us to  a number of   
results discussed in this mini-review and 
to  A. V. Kotikov for his contribution to our common works.   
It is a 
pleasure to thank S.I. Alekhin, G. Altarelli, S. Catani, B.I.  Ermolaev,  
S.A. Kulagin and 
F. J. Yndurain for many useful discussions. 
It is an honour to express my  warm gratitude to J. Tran Thanh Van, 
who contributed a lot 
to organizing   non-formal discussions between experimentalists and theoreticians 
during the Rencontres de Moriond QCD sessions, which had an 
essential influence on 
the works discussed above, and especially on  those   
devoted to the study of  neutrino DIS data  of the   CCFR collaboration. 
Special thanks go to M. Greco for giving me the possibility  to present the
talk at the  very productive La Thuile Conference.
I also would like to express my special  thanks 
to the members of the Theoretical Physics 
Division of CERN for creating a  pleasant  scientific atmosphere.

\end{document}